\documentstyle[preprint,revtex]{aps}	
\begin{document}
\topmargin 0.7in
\oddsidemargin 0.8in
\evensidemargin 0.8in
\draft
\hyphenation{brems-strahl-ung}
%
%
\preprint{FSU-HEP-940214}
\preprint{February 1994}
\begin{title}
Rapidity Correlations in $W\gamma$ Production\\
at Hadron Colliders
\end{title}
\author{U.~Baur}
\begin{instit}
Department of Physics, Florida State University, Tallahassee, FL 32306
\end{instit}
\vskip -2.mm
\author{S.~Errede}
\begin{instit}
Physics Department, University of Illinois, Urbana, IL 61801
\end{instit}
\vskip -2.mm
\author{G.~Landsberg}
\begin{instit}
Physics Department, SUNY at Stony Brook, NY 11794
\end{instit}
\thispagestyle{empty}
\vskip -15.mm
\begin{abstract}
\nonum\section{abstract}
\baselineskip16.pt  
We study the correlation of photon and charged lepton pseudorapidities,
$\eta(\gamma)$ and $\eta(\ell)$, $\ell=e,\,\mu$, in $p\,p\hskip-7pt
\hbox{$^{^{(\!-\!)}}$}\rightarrow W^\pm\gamma+X\rightarrow \ell^\pm
p\llap/_T\gamma+X$. In the Standard Model, the $\Delta\eta(\gamma,\ell)=
\eta(\gamma) - \eta(\ell)$
differential cross section is found to exhibit a pronounced dip at
$\Delta\eta(\gamma,\ell) \approx \mp 0.3$ ($=0$)in $p\bar p$ ($pp$)
collisions, which originates from the
radiation zero present in $q\bar q'\rightarrow W\gamma$. The sensitivity
of the $\Delta\eta(\gamma,\ell)$ distribution to higher order QCD
corrections, non-standard $WW\gamma$ couplings, the $W+$~jet ``fake''
background and the cuts imposed is
explored. At hadron supercolliders, next-to-leading order QCD corrections are
found to considerably obscure the radiation zero. The advantages of the
$\Delta\eta(\gamma,\ell)$ distribution over other quantities which are
sensitive to the radiation zero are discussed.
We conclude that photon lepton rapidity correlations at the
Tevatron offer a {\it unique} opportunity to search for the Standard Model
radiation zero in hadronic $W\gamma$ production.
\end{abstract}
\newpage
%
%
\begin{narrowtext}
\section{INTRODUCTION}

The electroweak Standard Model (SM) based on an
$\hbox{\rm SU(2)} \bigotimes \hbox{\rm U(1)}$ gauge theory
has been remarkably successful in describing contemporary high
energy physics experiments. The three vector boson couplings
predicted by this non-abelian gauge theory, however, remain largely untested.
The production of $W\gamma$ pairs at hadron colliders provides an
excellent opportunity to study the $WW\gamma$ vertex~\cite{BROWN,EHLQ}.
A pronounced feature of $W\gamma$ production in hadronic collisions is
the so-called radiation zero which appears in the parton level
subprocesses which contribute to lowest order in the SM of
electroweak interactions~\cite{RAZ}. For $u\bar d\to W^+\gamma$ ($d\bar u\to
W^-\gamma$) all contributing
helicity amplitudes vanish for $\cos\Theta^*=-1/3$ (+1/3), where
$\Theta^*$ is the angle between the quark and the photon in the parton
center of mass frame.

In practice, however, this zero is difficult to observe.
Structure function effects transform the zero into a dip in the
$\cos\Theta^*$ distribution~\cite{CHH}. Higher order QCD
corrections~\cite{NLO,NLOTWO}, and finite $W$ width effects, together with
photon radiation from the final state lepton line, tend to fill in this
dip~\cite{VALENZUELA}. Finally, finite detector resolution effects and
ambiguities in reconstructing the parton center of mass frame represent
an additional major complication in the extraction of the $\cos\Theta^*$ or
the corresponding rapidity distribution, $d\sigma/dy^*(\gamma)$, which
further dilute the signal of the radiation zero. The ambiguities are
associated with the
nonobservation of the neutrino arising from $W$ decay. Identifying the
missing transverse momentum with the transverse momentum of the neutrino
of a given $W\gamma$ event, the unobservable longitudinal neutrino
momentum, $p_L(\nu)$, and thus the parton center of
mass frame, can be reconstructed by imposing the constraint that the
neutrino and charged lepton four momenta combine to form the $W$ rest
mass~\cite{STROUGHAIR}. The resulting quadratic equation, in general, has
two solutions. In the approximation of a zero $W$ decay width, one of
the two solutions coincides with the true $p_L(\nu)$. On an event to
event basis, however, it is impossible to tell which of the two
solutions is the correct one. This ambiguity
considerably smears out the dip caused by the SM
radiation zero. Moreover, finite $W$ width effects in general cause a slight
mismatch between the true $p_L(\nu)$ and the correct solution for the
reconstructed longitudinal neutrino momentum which reduces the
significance of the dip even further.

The negative effect of higher order QCD corrections on the
observability of the SM radiation zero can partly be compensated by
imposing a jet veto~\cite{NLOTWO}. Event mis-reconstruction
originating from the twofold ambiguity in $p_L(\nu)$ can be
avoided by considering quantities which reflect the radiation zero, but
which do not require the reconstruction of the parton center of mass
frame. Recently, the ratio of $Z\gamma$ and $W^\pm\gamma$ cross sections as a
function of the minimum photon $p_T$ has been demonstrated~\cite{BEO}
to be such a quantity. However, a sufficiently large sample
of $Z\gamma$ events is necessary to establish the increase of the cross
section ratio with the minimum transverse momentum of the photon as
predicted by the SM. In
addition one has to assume that there are no non-standard contributions
to $Z\gamma$ production, {\it e.g.} from anomalous $ZZ\gamma$ or
$Z\gamma\gamma$ couplings~\cite{ELB}.

In this paper we consider correlations between the photon
pseudorapidity, $\eta(\gamma)$, and the pseudorapidity $\eta(\ell)$ of
the charged
lepton, $\ell=e,\,\mu$, originating from the $W$ decay as tools to
observe the radiation zero in $p\,p\hskip-7pt\hbox{$^{^{(\!-\!)
}}$}\rightarrow W\gamma$ predicted by the SM. In Section~II we study
the rapidity correlations at the Tevatron. We show that the double differential
distribution $d^2\sigma/d\eta(\gamma)d\eta(\ell)$ and the distribution of
the difference of pseudorapidities, $\Delta\eta(\gamma,\ell)=\eta(\gamma) -
\eta(\ell)$, clearly display the SM radiation zero. These distributions
do not require knowledge of the longitudinal momentum of the neutrino
and so automatically avoid the event mis-reconstruction problems
originating from the
two possible solutions for $p_L(\nu)$ which plague the $\cos\Theta^*$ and
$y^*(\gamma)$ distribution. In Section~II, we also study the
sensitivity of the $\Delta\eta(\gamma,\ell)$ spectrum to higher order QCD
corrections, non-standard $WW\gamma$ couplings, the $W+$~jet ``fake''
background and the cuts imposed. In
particular we find that the observability of the radiation zero strongly
depends on the photon lepton separation cut imposed and a sufficient
pseudorapidity coverage.

In Section~III we consider photon lepton rapidity correlations at future
$pp$ colliders. At LHC energies, $\sqrt{s}=14$~TeV, QCD corrections to
$W\gamma$ production are very large and obscure the effects caused by
the radiation zero. Even if a jet veto is imposed, it will not be easy
to observe the remnants of the radiation zero. For $pp$ collisions at
$\sqrt{s}=40$~TeV, the QCD corrections make it impossible to observe any
trace of the radiation zero in the $\Delta\eta(\gamma,\ell)$
distribution. Our conclusions are given in Section~IV. Some preliminary
results of the work described here were presented in Ref.~\cite{BEL}.

\section{PHOTON LEPTON RAPIDITY CORRELATIONS AT THE TEVATRON}

\subsection{Preliminaries}

In our analysis we shall focus entirely on the $W^+\gamma$ channel.
The rates for $W^+\gamma$ and $W^-\gamma$ production are equal in $p\bar p$
collisions. For $W^-\gamma$ production results therefore can be
obtained by exchanging the sign of the rapidities involved.
The calculation of $W\gamma$ production
in the Born approximation and at ${\cal O}(\alpha_s)$ is
performed using the results of Ref.~\cite{BB} and~\cite{NLOTWO},
respectively. In the Born approximation, our calculation fully
incorporates finite $W$
width effects, together with photon bremsstrahlung from the final state
charged lepton line (radiative $W$ decay). The next-to-leading order
(NLO) QCD calculation, on the other hand, treats the $W$ boson in the
narrow width approximation. In this approximation, diagrams in which the photon
is radiated off the final state lepton line are not necessary to maintain
electromagnetic gauge invariance. Imposing a sufficiently large photon lepton
separation cut, together with a cluster transverse mass cut, these
diagrams can be ignored~\cite{NLOTWO} (see below). In all our calculations,
we neglect the mass of the charged leptons so that the lepton rapidity,
\begin{eqnarray}
\noalign{\vskip 5pt}
y(\ell)={1\over 2}\log\left(E(\ell)+p_L(\ell)\over E(\ell)-p_L(\ell)\right)~,
\label{EQ:RAP}
\end{eqnarray}
and pseudorapidity,
\begin{eqnarray}
\eta(\ell)= -\log\left[\tan\left(\theta\over 2\right)\right]~,
\label{EQ:PRAP}
\end{eqnarray}
coincide. Here, $E(\ell)$ and $p_L(\ell)$ are the energy and the
longitudinal momentum, and $\theta$ is the scattering angle of the
lepton in the laboratory frame.
We assume that the incoming proton moves in the $+z$ direction.
The pseudorapidity has the advantage of being directly measurable
experimentally.

The signal in $p\bar p\rightarrow W^+\gamma$ consists of an
isolated high transverse momentum ($p_T$) photon and a $W^+$
boson which may decay either hadronically or leptonically. The hadronic
$W$ decays will be difficult to observe due to the large QCD
2~jet + $\gamma$ background~\cite{JJG}. In the following we therefore
focus on the leptonic $W$ decay modes. The signal for
$W^+\gamma$ production is then
\begin{equation}
p\bar p\rightarrow\ell^+ p\llap/_T\gamma+X,
\label{EQ:SIG}
\end{equation}
where $\ell=e,\,\mu$ (we neglect the $\tau$ decay mode of the $W$) and
the missing transverse momentum $p\llap/_T$ results from the
nonobservation of the neutrino from the $W$ decay.

In order to simulate the finite acceptance of detectors we impose,
unless stated otherwise, the following set of generic transverse momentum,
pseudorapidity, and separation cuts:
\begin{eqnarray}
p_T(\gamma)> 5~{\rm GeV}, & \qquad & |\eta(\gamma)|<3,  \nonumber\\
p_T(\ell)> 20~{\rm GeV}, & \qquad & |\eta(\ell)|<3.5, \label{EQ:TEVCUTS}
\\
p\llap/_T > 20~{\rm GeV},  &  \qquad &  \Delta R(\gamma,\ell) > 0.7.
\nonumber
\end{eqnarray}
Here,
\begin{equation}
 \Delta R(\gamma,\ell)=\left[\left(\Delta\Phi(\gamma,\ell)\right)^2 +
\left(\Delta\eta(\gamma,\ell)\right)^2\right]^{1/2}
\label{EQ:DELR}
\end{equation}
is the photon lepton separation in the pseudorapidity azimuthal
angle plane. The cuts listed in Eq.~(\ref{EQ:TEVCUTS}) approximate the
phase space region covered by the CDF and D\O\ detectors at the
Tevatron~\cite{CDF,DNULL}. Uncertainties in the energy measurements of
the final state particles in the detector are simulated by Gaussian
smearing of the particle four-momentum vector with standard deviation
$\sigma$, corresponding to the CDF detector resolution~\cite{CDF}. The
overall resolution of the D\O\ detector is comparable to that of
CDF~\cite{DNULL}.

Although we require a large lepton photon separation, radiative $W$
decays, $W\rightarrow\ell\nu\gamma$, still represent the majority of
$\ell p\llap/_T\gamma$ events, unless further cuts are imposed. The
contribution from radiative $W$ decay can be almost completely
eliminated by requiring
\begin{eqnarray}
M_T^{}(\ell \gamma;p\llap/_T^{}) > 90\ \hbox{\rm GeV} \>,
\label{EQ:MTCUT}
\end{eqnarray}
where
\FL
\begin{eqnarray}
M_T^2 (\ell \gamma;p\llap/_T^{}) =
\biggl[ \Bigl( M_{\ell \gamma}^2
 + \bigl| \hbox{\bf p}_{T}^{}(\gamma)
 + \hbox{\bf p}_{T}^{}(\ell) \bigr|^2 \Bigr)^{1/2}
 + p\llap/_T^{} \biggr]^2
- \bigl|   \hbox{\bf p}_{T}^{}(\gamma)
         + \hbox{\bf p}_{T}^{}(\ell)
         + \hbox{\bf p}\llap/_T^{} \bigr|^2 \, \> \>
\label{EQ:MTDEF}
\end{eqnarray}
is the square of the cluster transverse mass (also called minimum
invariant mass). In Eq.~(\ref{EQ:MTDEF}),
$M_{\ell \gamma}$ denotes the invariant mass of the $\ell \gamma$ system.
For $W \rightarrow\ell\nu\gamma$ the cluster transverse mass peaks sharply
at $M_W^{}$ (Ref.~\cite{CTMASS}) and drops rapidly above the $W$ mass.

The SM parameters used in our calculations are $\alpha=\alpha(M_Z^2)
=1/128$, $M_Z=91.18$~GeV, and $\sin^2\theta_W=0.23$. These values are
consistent with recent measurements at LEP, the CERN $p\bar p$
collider, and the Tevatron \cite{LEP,MW}.
For the parton distribution functions we use
the MRSS0 set~\cite{MRS} with the scale $Q^2$ given by the parton
center of mass energy squared, $\hat s$. The MRSS0 parametrization takes
into account the most recent NMC~\cite{NMC} and CCFR~\cite{CCFR} data,
and is compatible with measurements of the proton structure functions at
HERA~\cite{HERA}. Other recent fits to the parton distribution functions
such as the CTEQ parametrizations~\cite{CTEQ} lead to very similar
results for the kinematical quantities considered here.

\subsection{Rapidity Correlations}

The SM radiation zero leads to a
pronounced dip in the photon rapidity distribution in the center of
mass frame~\cite{BB}, $d\sigma/dy^*(\gamma)$, at
\begin{eqnarray}
y^*(\gamma)=y_0=-{1\over 2}\,\log 2\approx -0.35.
\label{EQ:YSTAR}
\end{eqnarray}
For $u\bar d\rightarrow W^+\gamma$ the photon and the $W$ are back to back
in the center of mass frame. The corresponding rapidity distribution of
the $W$ in the parton center of mass frame, $d\sigma/dy^*(W)$, thus also
exhibits a dip which is located at
\begin{eqnarray}
y^*(W)=y_1\approx 0.05.
\label{EQ:YSTARW}
\end{eqnarray}
If $W$ mass effects could be ignored, one would expect
the minimum to occur at $y^*(W)= -y_0$.
For $u\bar d\rightarrow W^+\gamma$, the photon and $W$
rapidities in the laboratory frame and in the parton center of mass
frame are related by
\begin{eqnarray}
y(\gamma) & = & {1\over 2}\log\left({x_1\over x_2}\right)+y^*(\gamma),
\\[2.mm]
y(W) & = & {1\over 2}\log\left({x_1\over x_2}\right)+y^*(W),
\end{eqnarray}
where $x_1$ and $x_2$ are the momentum fractions carried by the incoming
partons. In the
double differential distribution of the rapidities in the laboratory
frame, $d^2\sigma/dy(\gamma)dy(W)$, one therefore expects a ``valley'' for
rapidities satisfying the relation
\begin{eqnarray}
y(\gamma)-y(W) = y^*(\gamma)-y^*(W) = y_0-y_1\approx -0.4.
\end{eqnarray}
The $d^2\sigma/dy(\gamma)dy(W)$ distribution for $W^+\gamma$
production at the Tevatron in the Born approximation is shown in
Fig.~\ref{FIG:ONE}. The ``valley'' in
the double differential distribution of the rapidities is clearly
visible. Unfortunately, in order to determine the $W$ rapidity, the
longitudinal momentum of the neutrino, $p_L(\nu)$, must be known. Event
mis-reconstruction originating from the twofold ambiguity in $p_L(\nu)
$~\cite{STROUGHAIR} will therefore blur the observation of the
valley in $d^2\sigma/dy(\gamma)dy(W)$. Consequently, the double differential
distribution of photon and $W$ rapidities does not offer any significant
advantage over $d\sigma/dy^*(\gamma)$.

In the SM, the dominant $W$ helicity in $W^+\gamma$ ($W^-\gamma$)
production is $\lambda_W = +1$ ($\lambda_W = -1$)~\cite{BBS}, implying that
the charged lepton tends to be emitted in the direction of the parent
$W$, and thus reflects most of its kinematic properties. The difference
in rapidity, $\Delta y(W,\ell)=y(W)-y(\ell)$,
between the $W$ boson and the charged lepton originating from the $W$ decay
is small on the average. This is demonstrated in Fig.~\ref{FIG:TWO}
which shows the $\Delta y(W,\ell)$ distribution for the cuts listed in
Eqs.~(\ref{EQ:TEVCUTS}) and~(\ref{EQ:MTCUT}) (solid
line). The distribution sharply peaks at $\Delta y(W,\ell)\approx 0.3$.
For approximately 98\% of all events, $|\Delta y(W,\ell)|<1$. The $W$
lepton rapidity difference distribution is slightly asymmetric with
respect to the peak position. This is due
to the lepton and missing transverse momentum cuts, which favor events
with small lepton rapidity. If the $p_T(\ell)$ and
$p\llap/_T$ cuts are removed, the tail of the $\Delta y(W,\ell)$ distribution
extends to significantly higher positive rapidity differences (dashed line)
with the peak position remaining essentially unchanged.

Since the charged lepton carries most of the information of the $W$ one
expects that the valley signaling the SM radiation zero in
$d^2\sigma/dy(\gamma)dy(W)$ also manifests itself in
the double differential distribution $d^2\sigma/d\eta(\gamma)
d\eta(\ell)$ of the photon and lepton pseudorapidities in the laboratory
frame, $\eta(\gamma)$ and $\eta(\ell)$. The minimum is expected to be
located approximately at
\begin{eqnarray}
\Delta\eta(\gamma,\ell)=\eta(\gamma)-\eta(\ell) \approx y_0-y_1+\langle
\Delta y(W,\ell)\rangle ,
\label{EQ:MASTER}
\end{eqnarray}
where $\langle \Delta y(W,\ell)\rangle\approx 0.3$ is the average $W$
lepton rapidity difference.
Figure~\ref{FIG:THREE} shows $d^2\sigma/d\eta(\gamma)d\eta(\ell^+)$ for
$p\bar p\rightarrow \ell^+p\llap/_T\gamma$ in the Born approximation
(Fig.~\ref{FIG:THREE}a),
together with the corresponding distribution for $p\bar p\rightarrow
\ell^+\ell^-\gamma$ (Fig.~\ref{FIG:THREE}b). Since no radiation zero
occurs in any of the helicity amplitudes contributing to $p\bar p\rightarrow
\ell^+\ell^-\gamma$, one does not expect a valley in $d^2\sigma/d\eta(
\gamma)d\eta(\ell^+)$ for the $\ell^+\ell^-\gamma$ final state. The double
differential cross section for $p\bar p\rightarrow\ell^+\ell^-\gamma$ in
Fig.~\ref{FIG:THREE}b is computed using the results of
Ref.~\cite{ELB}, imposing the photon and lepton cuts of
Eq.~(\ref{EQ:TEVCUTS}). The calculation of Ref.~\cite{ELB} takes into
account the full set of Feynman diagrams contributing to $p\bar
p\rightarrow\ell^+\ell^-\gamma$, including radiative $Z$ decay and
timelike photon exchange graphs. In order to suppress contributions from
$Z\rightarrow\ell^+\ell^-\gamma$, the invariant mass of the lepton pair
and the $\ell\ell\gamma$ system are required to be
\begin{eqnarray}
m(\ell^+\ell^-)>70~{\rm GeV,} &\qquad & m(\ell^+\ell^-\gamma)>100~{\rm
GeV}.
\label{EQ:ZCUTS}
\end{eqnarray}

Figure~\ref{FIG:THREE}a indeed exhibits a pronounced minimum for
pseudorapidities satisfying $\Delta\eta(\gamma,
\ell)\approx -0.3$, which is only 0.2 units away from what one naively
expects (see Eq.~(\ref{EQ:MASTER})). The overall shape
of the double differential distribution of the photon and lepton
pseudorapidities and $d^2\sigma/dy(\gamma)dy(W^+)$ are very similar
(see Fig.~\ref{FIG:ONE}). In contrast to $d\sigma/dy^*(\gamma)$ and
$d^2\sigma/dy(\gamma)dy(W^+)$,
the double differential distribution of photon and lepton
pseudorapidities does not require knowledge of the longitudinal momentum
of the neutrino. Complications originating from the two possible
solutions for $p_L(\nu)$~\cite{STROUGHAIR} are therefore automatically
avoided.

Since the $W^+\gamma$ cross section is dominated by the contributions of the
$\lambda_W=+1$ helicity amplitudes, the photon and lepton
pseudorapidities are strongly correlated, with most photons
(leptons) having positive (negative) rapidities. For $W^-\gamma$
production, the $\lambda_W=-1$ amplitudes prevail, and
the sign of the rapidities changes. The correlation of
photon and lepton rapidities thus may aid
in determining the charge of the electron in $W\gamma$ events in the D\O \
detector which does not have a central magnetic field~\cite{DNULL}. Without the
cluster transverse mass cut of $M_T(\ell\gamma;p\llap/_T)>90$~GeV,
radiative $W$ decays dominate and the valley disappears. The
``channel'' in $d^2\sigma/d\eta(\gamma)d\eta(\ell^+)$ is
approximately 1.5 units in rapidity wide and occurs
essentially in the diagonal of the $\eta(\gamma),\eta(\ell)$ plane. It
is therefore completely obscured if one integrates over the full range
of either the photon or the lepton pseudorapidity.
Only for sufficiently strong cuts, {\it e.g.} $|\eta(\gamma)|<1$ or
$|\eta(\ell)|<1$, a slight dip can be observed in
$d\sigma/d\eta(\ell)$ ($d\sigma/d\eta(\gamma)$) in the region around
$\eta(\ell)\approx 0.3$ ($\eta(\gamma)\approx -0.3$).

In contrast to the situation
for $p\bar p\rightarrow\ell^+p\llap/_T\gamma$, there is no sign of a
valley, and no strong correlation between the rapidities in $p\bar
p\rightarrow\ell^+\ell^-\gamma$ (see Fig.~\ref{FIG:THREE}b). The
slightly asymmetric
double peak structure originates from the collinear singularity
associated with the contribution from radiative $Z$ decays and the
$\Delta R(\gamma,\ell)>0.7$ cut. For the cuts
imposed, approximately 25\% of the cross section originates from
$Z\rightarrow\ell^+\ell^-\gamma$ decays. For $|\Delta\eta(\gamma,\ell^+)
|< 0.7$ the $\Delta R(\gamma,\ell)$ cut enforces a large azimuthal
angle between the photon and the $\ell^+$ which reduces the cross
section in this region. The asymmetry can be traced to the interference of
the radiative $Z$ decay diagrams with the corresponding Feynman graphs
where the $Z$ boson is replaced by timelike virtual photon exchange.

The double differential distribution $d^2\sigma/d\eta(\gamma)
d\eta(\ell^+)$ can only be mapped out if a sufficiently large number of
events is available. For a relatively small $W\gamma$ event sample, the
distribution of the
rapidity difference, $d\sigma/d\Delta\eta(\gamma,\ell^+)$, is more
useful. The $\Delta\eta(\gamma,\ell^+)$ differential cross section for $p\bar
p\rightarrow\ell^+p\llap/_T\gamma$ and $p\bar
p\rightarrow\ell^+\ell^-\gamma$ and the cuts summarized in
Eqs.~(\ref{EQ:TEVCUTS}), (\ref{EQ:MTCUT}) and (\ref{EQ:ZCUTS}) is shown in
Fig.~\ref{FIG:FOUR} by the solid and dashed line, respectively.
The SM radiation zero leads to a strong dip in the
$\Delta\eta(\gamma,\ell^+)$ distribution for $p\bar p\rightarrow
\ell^+p\llap/_T\gamma$ at $\Delta\eta(\gamma,\ell)\approx -0.3$. For
$p\bar p\rightarrow\ell^+\ell^-\gamma$, on the other
hand, $d\sigma/d\Delta\eta(\gamma,\ell^+)$ shows a slightly
asymmetric double peak structure, which, as explained above, is caused
by the collinear divergence of the final state bremsstrahlung diagrams
together with the finite $\Delta R(\gamma,\ell)$ cut and the
interference of $Z$ and photon exchange diagrams. Since the rapidity
difference distribution can be analyzed more easily than
$d^2\sigma/d\eta(\gamma)d\eta(\ell)$, we shall focus on
$d\sigma/d\Delta\eta(\gamma,\ell)$ in the following.

\subsection{Properties of the Pseudorapidity Difference Distribution}

We now study the properties of the pseudorapidity difference
distribution in more detail.
The large asymmetry in the $\Delta\eta(\gamma,\ell)$ differential
cross section between positive and negative values of $\Delta\eta(\gamma,
\ell)$ originates from valence quark effects, combined with
the difference in the electric charges of $u$ and $d$-quarks. This is
illustrated in more detail in Fig.~\ref{FIG:FIVE}, where we show the individual
contributions of valence -- valence, sea -- sea and valence -- sea
quark collisions to the $\Delta\eta(\gamma,
\ell)$ distribution. The dotted line displays the pseudorapidity
difference distribution for sea quarks colliding with sea quarks.
The distribution is symmetric with respect to $\Delta\eta(\gamma,
\ell)=0$, and exhibits a pronounced minimum for zero pseudorapidity
differences. The dashed curve shows the distribution assuming that only
the valence quarks in the proton and antiproton contribute, whereas the
dash-dotted line displays $d\sigma/d\Delta\eta(\gamma,\ell)$ for valence
quarks colliding with sea quarks. Both curves exhibit a strong
asymmetry. In each case we have summed over all contributing
combinations and quark flavors. The solid line finally
gives the full pseudorapidity difference distribution. Note that
the sea quark cross section (dotted line) represents the single most
important contribution for $\Delta\eta(\gamma,\ell)<0$, whereas it is
by far the smallest one for positive pseudorapidity differences.
The asymmetry in the contributions involving valence quarks originates
from the Feynman graphs where the photon is emitted by the incoming $u$
or $\bar d$ quark (initial state radiation). Due to the collinear
singularity associated with these diagrams, photons radiated off the
incoming $u$ ($\bar d$) valence quark tend to be emitted in the
direction of the incident proton (antiproton) beam, thus having mostly
$\Delta\eta(\gamma,\ell)>0$ ($\Delta\eta(\gamma,\ell)<0$).

In order to assess the observability of the radiation zero in the
$\Delta\eta(\gamma,\ell)$ distribution, it is important to know how
sensitive the dip in
$d\sigma/d\Delta\eta(\gamma,\ell)$ is to the cuts imposed. The variation
of the $\Delta\eta(\gamma,\ell)$ distribution with the geometrical
acceptances is explored in Figs.~\ref{FIG:SIX} and~\ref{FIG:SEVEN}.
Whereas Fig.~\ref{FIG:SIX} presents a global view of the pseudorapidity
difference spectrum, Fig.~\ref{FIG:SEVEN} focuses
on the dip region to present a more detailed view of the region of
interest. In the vicinity of the dip, most events originate from the high
$p_T(\gamma)$ region. This is illustrated by the dashed line in
Fig.~\ref{FIG:SIX}a and~\ref{FIG:SEVEN},
which shows the $\Delta\eta(\gamma,\ell)$ distribution for $p_T(\gamma)
>10$~GeV instead of $p_T(\gamma)>5$~GeV. Around the
minimum, the result almost coincides with that obtained for the smaller
photon $p_T$ cut; see Fig.~\ref{FIG:SEVEN} (dashed line).
Increasing the photon transverse momentum threshold, the dip clearly
becomes less pronounced. In contrast to the photon transverse momentum
cut, the $p_T(\ell)$ and $p\llap/_T$ requirements only marginally influence the
depth of the dip. This is demonstrated by the dash-dotted lines in
Figs.~\ref{FIG:SIX}a and~\ref{FIG:SEVEN}, which show the
$\Delta\eta(\gamma,\ell)$ distribution if the lepton and missing $p_T$
cuts are removed.

The sensitivity of $d\sigma/d\Delta\eta(\gamma,\ell)$ to the $\Delta
R(\gamma,\ell)$ cut imposed is explored in Fig.~\ref{FIG:SIX}b where the
pseudorapidity difference distribution obtained with the cuts of
Eqs.~(\ref{EQ:TEVCUTS}) and~(\ref{EQ:MTCUT}) (solid line) is compared with
the result obtained if the $\Delta R(\gamma,\ell)>0.7$ requirement
is replaced by $\Delta R(\gamma,\ell)>0.3$, with all
other cuts unchanged (dotted line). Reducing the photon
lepton separation cut increases the contribution of the diagram
where the photon is radiated off the final state lepton line. The final
state bremsstrahlung contribution, which diverges in the collinear
limit, destroys the SM radiation zero, and thus tends to fill in the
dip. Closer inspection of the region around the minimum, however, reveals an
interesting pattern.
The dip is not filled in uniformly (see Fig.~\ref{FIG:SEVEN}). For values of
$|\Delta\eta(\gamma,\ell)|$ smaller than the $\Delta R(\gamma,\ell)$
cut, a large azimuthal angle between the photon and the lepton is
enforced, resulting in a ``dent'' in the
distribution. If the photon lepton separation cut is further reduced,
the collinear singularity of the final state lepton bremsstrahlung leads
to a strong peak at $\Delta\eta(\gamma,\ell)=0$. This is demonstrated
by the long-dashed line in Fig.~\ref{FIG:SEVEN}, which shows
$d\sigma/d\Delta\eta(\gamma,\ell)$ for $\Delta R(\gamma,\ell)>0.1$, and
all other cuts as in Eqs.~(\ref{EQ:TEVCUTS}) and~(\ref{EQ:MTCUT}). The
characteristic dent caused by the photon lepton separation cut is again
clearly visible. Whereas the dip caused by the radiation zero
is filled in if the $\Delta R(\gamma,\ell)$ cut is reduced,
imposing a photon lepton separation cut larger than that of
Eq.~(\ref{EQ:TEVCUTS}), does not improve the significance of the radiation
zero noticeably.

For a photon and/or lepton pseudorapidity range $|\eta(\gamma)
|<\eta_{\rm max}(\gamma)$, $|\eta(\ell)|<\eta_{\rm max}(\ell)$,
smaller than that of Eq.~(\ref{EQ:TEVCUTS}), the $\Delta\eta(\gamma,\ell)$
distribution is truncated at
$|\Delta\eta(\gamma,\ell)|=\eta_{\rm max}(\gamma)+\eta_{\rm max}(\ell)$.
Details depend on the specific values of $\eta_{\rm max}(\gamma)$ and
$\eta_{\rm max}(\ell)$.
If photons and leptons are both simultaneously restricted to the
central rapidity region, $|\eta(\gamma)|$, $|\eta(\ell)|<1$, the event
rate and the significance of the minimum at $\Delta\eta(\gamma,\ell)
\approx -0.3$ is reduced substantially (dashed
double dotted line in Fig.~\ref{FIG:SIX}b). In this
case one only selects events from the region around the minimum in
$d^2\sigma/d\eta(\gamma)d\eta(\ell)$ (see Fig.~\ref{FIG:THREE}) and the
$\Delta\eta(\gamma,\ell)$ distribution largely looses its usefulness.

So far, our calculations have been carried out entirely in the Born
approximation. Higher order QCD corrections are known to eliminate the
radiation zero~\cite{NLO,NLOTWO}. It is therefore necessary to study
the $\Delta\eta(\gamma,\ell)$ distribution at ${\cal O}(\alpha_s)$.
The pseudorapidity difference distribution at next-to-leading order in
QCD is shown in Fig.~\ref{FIG:EIGHT}. Besides the cuts described above, we also
require the photon to be isolated from hadrons in the NLO calculation by
imposing a cut on the total additional hadronic energy in a cone of
size $\Delta R=0.7$ about the direction of the photon of~\cite{EPS}
\begin{eqnarray}
\sum_{\Delta R < 0.7} \, E_{\hbox{\scriptsize had}} < 0.15 \, E_{\gamma} \>,
\label{EQ:ISOL}
\end{eqnarray}
where $E_\gamma$ is the photon energy. This requirement strongly reduces
photon bremsstrahlung from final state quarks and gluons.

Figure~\ref{FIG:EIGHT}a displays the inclusive NLO ($W\gamma+X$),
the leading order (LO) $W\gamma+1$~jet, and the Born $\Delta\eta(\gamma,\ell)$
distribution. In Fig.~\ref{FIG:EIGHT}b the exclusive NLO
$W\gamma+0$~jet and Born distributions are compared. Here, a jet is
defined as a quark or gluon with
\begin{eqnarray}
p_T^{}(j)>10~{\rm GeV}\hskip 1.cm {\rm and} \hskip 1.cm |\eta(j)|<2.5.
\label{EQ:TEVJET}
\end{eqnarray}
As mentioned at the beginning of this
section, our NLO calculation of $W\gamma$ production is carried out
using the narrow width approximation for the $W$ decay and ignoring
bremsstrahlung from the final state lepton line. In order to perform a
meaningful comparison, we have calculated the Born cross section in
Fig.~\ref{FIG:EIGHT} using the same approximations. The
$\Delta\eta(\gamma,\ell)$
distribution in the Born approximation so obtained is very similar to
that found calculating the full set of Feynman diagrams for the cuts we
impose (see Eqs.~(\ref{EQ:TEVCUTS}) and~(\ref{EQ:MTCUT})), except the immediate
region around the minimum at $\Delta\eta(\gamma,\ell)\approx -0.3$, where the
approximation neglecting bremsstrahlung from the final state lepton line
leads to a more pronounced dip, as expected.

NLO QCD corrections are seen to partially
fill in the dip caused by the SM radiation zero. While
${\cal O}(\alpha_s)$ QCD corrections enhance the cross section by 30 --
40\% outside the dip region, they increase the rate by approximately a
factor~2 for $\Delta\eta(\gamma,\ell)\approx -0.3$ (see
Fig.~\ref{FIG:EIGHT}a). This effect is
predominantly caused by events containing a hard jet (dashed line).
However, next-to-leading log QCD corrections do not significantly obscure
the radiation zero. Imposing a jet veto, {\it
i.e.} requiring that no jets with transverse momentum $p_T(j)>10$~GeV and
rapidity $|\eta(j)|<2.5$ are present in the event (see
Fig.~\ref{FIG:EIGHT}b), the NLO
$\Delta\eta(\gamma,\ell)$ distribution (dotted line) is very
similar to that obtained in the Born approximation (dash-dotted line).
For jet definition criteria different from those of Eq.~(\ref{EQ:TEVJET}), the
agreement between the NLO 0-jet and the Born $\Delta\eta(\gamma,\ell)$
distribution is slightly worse. In general, for a $p_T(j)$ threshold
larger (smaller) than that of Eq.~(\ref{EQ:TEVJET}), the ${\cal
O}(\alpha_s)$ 0-jet $\Delta\eta(\gamma,\ell)$ differential cross
section will be somewhat above (below) the
prediction obtained in the Born approximation. Note that the jet
transverse momentum threshold cannot be lowered to
arbitrarily small values in our calculation for theoretical reasons.
For transverse momenta below 5~GeV, soft gluon
resummation effects are expected to significantly change
the jet $p_T^{}$ distribution~\cite{RESUM} at the Tevatron. These
effects are not included in our calculation.

At Tevatron energies, the dominant
contributions to the NLO cross section originate from quark-antiquark
annihilation.
Apart from the photon bremsstrahlung contribution, which is strongly
suppressed by the photon isolation cut, Eq.~(\ref{EQ:ISOL}),
all $2\rightarrow 2$ terms
are proportional to the $q\bar q'\rightarrow W\gamma$ matrix element
in the Born approximation and, therefore, preserve the radiation zero.
Furthermore, the $2\rightarrow 3$ process $q\bar q'\rightarrow
W^\pm\gamma g$ exhibits a radiation zero at $\cos\Theta^*=\mp 1/3$ if
the gluon is
collinear with the photon~\cite{BRO}, and also in the soft gluon limit,
$E_g\rightarrow 0$. The remnants of these zeros are responsible for the
slight depression observed in the LO 1-jet $\Delta\eta(\gamma,\ell)$
distribution (dashed line) at $\Delta\eta(\gamma,\ell)\approx -0.3$. In
the 0-jet configuration, contributions from the $2\rightarrow 3$
processes are suppressed. Since all $2\rightarrow 2$ contributions
except the photon bremsstrahlung term, which contributes negligibly for
the photon isolation cut we impose, preserve the radiation zero, it
is not surprising that the NLO 0-jet and Born $\Delta\eta(\gamma,\ell)$
distributions are very similar.

Besides bremsstrahlung contributions from the final state lepton line
and higher order QCD corrections, non-standard $WW\gamma$ couplings also
destroy the radiation zero. In $W\gamma$ production both the virtual $W$ and
the
decaying onshell $W$ couple to essentially massless fermions, which
insures that effectively $\partial_\mu W^\mu=0$. This condition
together with Lorentz invariance, electromagnetic gauge invariance, and $CP$
conservation, allows two free parameters, $\kappa$ and $\lambda$,
in the $WW\gamma$ vertex. The most general vertex compatible with these
conservation laws is described by the effective Lagrangian~\cite{LAGRANGIAN}
\begin{eqnarray}
\noalign{\vskip 5pt}
{\cal L}_{WW\gamma} &=& -i \, e \,
\Biggl[ W_{\mu\nu}^{\dagger} W^{\mu} A^{\nu}
              -W_{\mu}^{\dagger} A_{\nu} W^{\mu\nu}
+ \kappa W_{\mu}^{\dagger} W_{\nu} F^{\mu\nu}
+ {\lambda \over M_W^2} W_{\lambda \mu}^{\dagger} W^{\mu}_{\nu} F^{\nu\lambda}
\Biggr] \>,
\label{EQ:LAGRANGE}
\end{eqnarray}
where $A^{\mu}$ and $W^{\mu}$ are the photon and $W^-$ fields, respectively,
$W_{\mu\nu} = \partial_{\mu}W_{\nu} - \partial_{\nu}W_{\mu}$, and
$F_{\mu\nu} = \partial_{\mu}A_{\nu} - \partial_{\nu}A_{\mu}$.
The variables $\kappa$ and $\lambda$ are in the static limit related to
the magnetic dipole moment, $\mu_W^{}$, and the electric quadrupole
moment, $Q_W^{}$, of the $W$-boson:
\begin{eqnarray}
\noalign{\vskip 5pt}
\mu_{W}^{} = {e \over 2 M_W^{} }\, (1 + \kappa + \lambda) \>, \hskip
1.cm
Q_{W}^{} = -{e \over M_W^2 } \, (\kappa - \lambda) \>.
\end{eqnarray}
At tree level in the SM, $\kappa = 1$ and $\lambda = 0$.

If $CP$ violating $WW\gamma$ couplings are allowed, two additional
free parameters, $\tilde\kappa$ and $\tilde\lambda$ appear in the
effective Lagrangian. However,
$CP$ violating operators are tightly constrained by measurements of the neutron
electric dipole moment which restrict $\tilde \kappa$ and $\tilde \lambda$ to
$|\tilde \kappa|$, $|\tilde \lambda| < {\cal
O}(10^{-3})$~\cite{CPLIMITS}. $CP$ violating $WW\gamma$ couplings are,
therefore, not considered here.

Tree level unitarity uniquely restricts the $WW\gamma$ couplings to their
SM gauge theory values at asymptotically high energies~\cite{CORNWALL}.
This implies that any deviation of $\kappa$ or $\lambda$ from the SM
expectation has to be described by a form factor
$a(M_{W\gamma}^2, p_W^2, p_\gamma^2)$, $a=(\Delta\kappa=\kappa-1),\,
\lambda$ which
vanishes at high energies. Consequently, the anomalous couplings are introduced
via form factors~\cite{WILLEN}
\begin{eqnarray}
\noalign{\vskip 2pt}
a(M_{W\gamma}^2, p_W^2 = M_W^2, p_\gamma^2 = 0) \> &=& \>
{a_0 \over (1 + M_{W\gamma}^2/\Lambda^2)^n } \>,
\label{EQ:KAPPAFORM}
\end{eqnarray}
where $a_0=\Delta \kappa_0,\,\lambda_0$ are the form factor values at
low energies and
$\Lambda$ represents the scale at which new physics becomes important in the
weak boson sector, {\it e.g.} due to a composite structure of the
$W$-boson. $M_{W\gamma}$ is the invariant mass of the $W\gamma$ system.
In order to guarantee unitarity, $n>1/2$ for $\Delta\kappa$ and $n>1$
for $\lambda$. For the numerical results presented below, we use a dipole
form factor ($n=2$) with a scale $\Lambda = 1$~TeV. The exponent $n=2$
is chosen in order to suppress $W\gamma$ production at energies
$\sqrt{\hat s}\gg\Lambda\gg M_W$, where novel phenomena like resonance
or multiple weak boson production are expected to become important.

The two $CP$ conserving couplings have recently been measured at the
CERN $p\bar p$ collider and at the Tevatron. The UA2
Collaboration at CERN obtained~\cite{UA2}
\newcommand{\crc}{\crcr\noalign{\vskip -8pt}}
\begin{eqnarray}
\noalign{\vskip 5pt}
\kappa_0=1\matrix{+2.6\crc -2.2}~~({\rm for}~\lambda_0=0) \>, \hskip 1.cm
\lambda_0=0\matrix{+1.7\crc -1.8}~~({\rm for}~\kappa_0=1) \>,
\label{EQ:UA2}
\end{eqnarray}
at the 68.3\% confidence level (CL) from a fit to the photon $p_T$
distribution in the process $p\bar p\rightarrow e^\pm\nu\gamma X$. CDF
and D\O\ have extracted 68.3\% CL limits from the total
$W\gamma$ cross section, combining data from $W\rightarrow e\nu$ and
$W\rightarrow\mu\nu$ decays~\cite{THOMAS,TONY}. From the
analysis of the 1988-89 data, CDF finds~\cite{THOMAS}
\begin{eqnarray}
\noalign{\vskip 5pt}
\kappa_0=1\matrix{+3.7\crc -3.2}~~({\rm for}~\lambda_0=0) \>, \hskip 1.cm
\lambda_0=0\matrix{+1.6\crc -1.6}~~({\rm for}~\kappa_0=1) \>,
\label{EQ:CDF}
\end{eqnarray}
whereas D\O\ obtains the following preliminary limits from data
taken during 1992-93 Tevatron run~\cite{TONY}:
\begin{eqnarray}
\noalign{\vskip 5pt}
\kappa_0=1\matrix{+1.8\crc -1.5}~~({\rm for}~\lambda_0=0) \>, \hskip 1.cm
\lambda_0=0\matrix{+0.71\crc -0.75}~~({\rm for}~\kappa_0=1) \>.
\label{EQ:DNULL}
\end{eqnarray}
CDF is expecting similar bounds from their analysis of the 1992-93 data.

Although bounds on these couplings can also be extracted from
low energy data and high precision measurements at the $Z$ pole, there are
ambiguities and model dependencies in the results~\cite{LOW}.
No rigorous bounds on $WW\gamma$ couplings can be
obtained from LEP~I data if correlations between different
contributions to the anomalous couplings are fully taken into account.

The sensitivity of the pseudorapidity difference distribution in the Born
approximation to non-standard $WW\gamma$ couplings is explored in
Fig.~\ref{FIG:NINE}.
In presence of any anomalous contribution to the $WW\gamma$ vertex the
radiation zero is eliminated and the dip in $d\sigma/d\Delta\eta(\gamma,
\ell)$ is filled in at least partially. This is illustrated by the
dashed and dotted curves, which show the result obtained for
the UA2 68.3\% CL limits of $\Delta\kappa_0=2.6$ and
$\lambda_0=1.7$ respectively. Only one anomalous coupling is varied at
a time here. Most of the excess cross section
for non-standard couplings originates in the high $p_T(\gamma)$ region
\cite{BB}, where photons tend to be central in rapidity. Deviations from
the SM $\Delta\eta(\gamma,\ell)$ distribution, therefore, mostly occur
for small pseudorapidity differences. In Fig.~\ref{FIG:NINE} we have
also included the  statistical uncertainties expected in the SM case
for an integrated luminosity of $\int\!{\cal L} dt=22$~pb$^{-1}$. This
demonstrates that the pseudorapidity
difference distribution is sensitive to anomalous $WW\gamma$ couplings
already with the current CDF and D\O \ data samples, in particular
to $\lambda$. However, we do not expect $d\sigma/d\Delta\eta(\gamma,
\ell)$ to be more sensitive to anomalous couplings than the photon
transverse momentum distribution.

The size of the error bars in Fig.~\ref{FIG:NINE}
indicates that it will probably not be possible to conclusively establish
the dip in the $\Delta\eta(\gamma,\ell)$ distribution with the small
number of events
CDF and D\O\ have presently available. In this situation the ratio
of cross sections for $\Delta\eta(\gamma,\ell)<0$ and
$\Delta\eta(\gamma,\ell)>0$,
\begin{eqnarray}
\noalign{\vskip 5pt}
R_{-+}={\int_{\Delta\eta(\gamma,\ell)<0}~{d\sigma\over d\Delta\eta(\gamma,
\ell)}\,~d\Delta\eta(\gamma,\ell)\over \int_{\Delta\eta(\gamma,\ell)>0}~
{d\sigma\over d\Delta\eta(\gamma,\ell)}\,~d\Delta\eta(\gamma,\ell)}~,
\label{EQ:RATIO}
\end{eqnarray}
may be useful to test whether the experimental data are consistent with the
SM prediction. Many experimental uncertainties, for example, those
associated with lepton and photon detection efficiencies, or the
uncertainty in the integrated luminosity, cancel completely in $R_{-+}$.
For the cuts of Eqs.~(\ref{EQ:TEVCUTS}) and~(\ref{EQ:MTCUT})
we find $R_{-+}=0.24$ in the SM, while we obtain $R_{-+}=0.37$ (0.47)
for $\Delta\kappa_0=2.6$ ($\lambda_0=1.7$). Alternatively,
$d\sigma/d\Delta\eta(\gamma,\ell)$ can be normalized to
$\sigma(W\rightarrow\ell\nu)$. In the resulting differential cross
section ratio, most experimental and theoretical uncertainties again cancel.
With an integrated luminosity of 100~pb$^{-1}$, which is
expected by the end of this year,
it should be possible to observe the dip in the photon lepton
pseudorapidity difference spectrum. A detailed mapping of the
$\Delta\eta(\gamma,\ell)$ distribution requires at least
$\sim 500$~pb$^{-1}$.

So far, we have completely ignored the possible effect of background
processes on
the rapidity difference distribution. The most serious background to
$W\gamma$ production in hadronic collisions originates from $W+$~jet
production with the jet misidentified as a photon. Such
misidentifications are mostly caused by jets hadronizing with a leading
$\pi^0$, which carries away most of the jet energy. The two photons
resulting from the $\pi^0$ decay may not be resolved in the detector.
The probability $P_{\gamma /j}$ that a jet ``fakes'' a photon is
detector specific and depends on the offline reconstruction method.
The photon
jet misidentification probabilities recently measured by UA2~\cite{UA2}
and CDF rapidly fall with increasing jet transverse energies. In
contrast, D\O\ finds $P_{\gamma /j}$ to be almost independent of the jet
$E_T$ in the region of interest~\cite{TONY}.

The rapidity difference distribution of the $W^++$~jet fake background,
using the central values of the CDF photon jet misidentification
probability and imposing the cuts of Eqs.~(\ref{EQ:TEVCUTS})
and~(\ref{EQ:MTCUT}), is shown by the solid line in Fig.~\ref{FIG:TEN}a.
For $p_T(\gamma)=5$~GeV , $P_{\gamma /j}={\cal O}(10^{-3})$, and
drops to $P_{\gamma /j}=~{\rm few}~\times 10^{-4}$ at $p_T(\gamma)=20$~GeV.
The dashed line gives the prediction for a constant misidentification
probability of $P_{\gamma /j}=1\cdot 10^{-3}$, which is consistent with
the value obtained by D\O~\cite{TONY}. For a realistic assessment of
the $W+$~jet
background, a full-fledged Monte Carlo simulation is required. Here, for
a first estimate of its impact on the shape of the pseudorapidity difference
distribution, we use a simple parton level calculation.
The experimental uncertainties in $P_{\gamma /j}$ are
presently quite large, resulting in a normalization uncertainty of the
solid curve in Fig.~\ref{FIG:TEN}a of about a factor~2.
However, the shape of the distribution changes only little if
$P_{\gamma /j}$ is varied within the allowed range.

The $\Delta\eta(\gamma,\ell)$ distribution of the $W+$~jet fake
background, using the CDF misidentification probability, displays a
very pronounced dip centered at
$\Delta\eta(\gamma,\ell)=0$, although none of the subprocesses which
contribute to $W+1$~jet production ($q\bar q'\rightarrow Wg$ and
$qg\rightarrow Wq'$) exhibits a radiation zero. The dip is primarily
caused by the strong variation of $P_{\gamma /j}$ with the transverse
energy. Because the misidentification probability drops rapidly with
$p_T$, the background cross section is significantly suppressed at high
transverse momenta. High $p_T$ events usually are very central in
rapidity, and thus cluster around $\Delta\eta(\gamma,\ell)=0$. The cross
section deficit at high transverse momenta thus leads to a dip in the
$\Delta\eta(\gamma,\ell)$ distribution, {\it i.e.} the $W+$~jet fake
background exhibits a ``fake radiation zero''. In contrast, the $W+$~jet
background obtained for a constant misidentification probability (dashed
line) shows only a slight dent, which is entirely due to the
$\Delta R(\gamma,\ell)>0.7$ cut imposed.

In Fig.~\ref{FIG:TEN}b we show the effect of the $W+$~jet fake
background on the $W\gamma$ rapidity difference distribution. The dotted
line shows the expected distribution of signal plus background, using the
($p_T$ dependent) misidentification probability of CDF, whereas the
dashed line shows the result for a constant $P_{\gamma /j}$ of
$P_{\gamma /j}=1\cdot 10^{-3}$. For comparison, we have also included
the prediction of the signal alone (solid line). The $W+$~jet background is
seen to mostly affect the $\Delta\eta(\gamma,\ell)$ distribution for
$\Delta\eta(\gamma,\ell)<0$. The significance of the dip is affected in
a noticeable way only in the case of a constant misidentification
probability.

\section{RAPIDITY CORRELATIONS AT HADRON SUPERCOLLIDERS}

In Section~II we presented a detailed analysis of photon lepton
pseudorapidity correlations at the Tevatron. We
now repeat the most important steps of this analysis for the planned
Large Hadron Collider (LHC) ($pp$
collisions at $\sqrt{s}=14$~TeV~\cite{LHC}). We also comment briefly on
what to expect at even higher energies. To simulate detector response,
we shall impose the following set of cuts,
\begin{eqnarray}
p_T(\gamma)> 100~{\rm GeV}, & \qquad & |\eta(\gamma)|<2.5,  \nonumber\\
p_T(\ell)> 25~{\rm GeV}, & \qquad & |\eta(\ell)|<3, \label{EQ:LHCCUTS}\\
p\llap/_T > 50~{\rm GeV},  &  \qquad &  \Delta R(\gamma,\ell) > 0.7,
\nonumber
\end{eqnarray}
together with the photon isolation requirement of Eq.~(\ref{EQ:ISOL}).
The large $p_T^{}(\gamma)$ and $p\llap/_T^{}$ cuts are chosen to
reduce potentially dangerous backgrounds from
$W+1$~jet production, where the jet is misidentified as a photon, and
from processes where particles outside the rapidity range covered by the
detector contribute to the missing transverse momentum. Present
studies~\cite{FAKE,ATLAS,CMS} indicate that these backgrounds are
under control for
$p_T^{}(\gamma) > 100$~GeV and $p\llap/_T^{}>50$~GeV. The large photon
transverse momentum cut automatically selects
a region of phase space in which $W\gamma$ production dominates. Energy
mismeasurements in the detector are simulated by Gaussian smearing of
the final state particle momenta corresponding to the resolution
foreseen for the ATLAS experiment~\cite{ATLAS}.

In the following we concentrate on the $\Delta\eta(\gamma,\ell)$
distribution for $pp\rightarrow W^+\gamma+X$ in the SM. For the cuts we
impose, the $W^-\gamma$ cross section at $pp$ supercolliders is slightly
smaller than that of $W^+\gamma$ production.
In $pp$ collisions $d\sigma/d\Delta\eta(\gamma,\ell)$ is
necessarily symmetric around $\Delta\eta(\gamma,\ell)=0$ and the
pseudorapidity difference distributions of $pp\rightarrow W^+\gamma+X$ and
$pp\rightarrow W^-\gamma+X$ are identical in shape. The dip
signaling the radiation zero is shifted to $\Delta\eta(\gamma,
\ell)=0$.

Due to the large $qg$ luminosity, the inclusive NLO QCD corrections
are very large for $W\gamma$ production at supercollider
energies~\cite{NLOTWO}. For $pp$ collisions at $\sqrt{s}=14$~TeV they
enhance the cross section by more than a factor~3.
This is demonstrated in Fig.~\ref{FIG:ELEVEN}, where we show the
photon lepton pseudorapidity difference distribution at the LHC for the
cuts summarized in Eq.~(\ref{EQ:LHCCUTS}). Figure~\ref{FIG:ELEVEN}a displays
the
inclusive NLO, the ${\cal O}(\alpha_s)$ 0-jet and the LO 1-jet
$\Delta\eta(\gamma,\ell)$ distributions.
The exclusive NLO 0-jet and the Born pseudorapidity difference
distributions are compared in Fig.~\ref{FIG:ELEVEN}b.
Here, a jet is defined as a quark or gluon with
\begin{eqnarray}
p_T^{}(j)>50~{\rm GeV}\hskip 1.cm {\rm and} \hskip 1.cm |\eta(j)|<3.
\label{EQ:LHCJET}
\end{eqnarray}
Whereas the Born $\Delta\eta(\gamma,\ell)$ distribution displays a
pronounced dip (dash-dotted curve in
Fig.~\ref{FIG:ELEVEN}b), the inclusive NLO pseudorapidity difference
distribution only
shows a slight depletion at zero rapidity difference (solid line in
Fig.~\ref{FIG:ELEVEN}a). In contrast to the situation encountered at the
Tevatron, QCD corrections thus considerably obscure the radiation zero
at supercollider energies.

The bulk of the QCD corrections to $W\gamma$ production at the LHC
originates from
quark gluon fusion and the kinematical region where the photon is
produced at large $p_T^{}$ and recoils against a quark, which radiates a
soft $W$ boson which is almost collinear to the quark~\cite{NLOTWO}.
Events which
originate from this phase space region usually contain a high $p_T$ jet.
The 1-jet $\Delta\eta(\gamma,\ell)$ distribution is given by the dashed
line in Fig.~\ref{FIG:ELEVEN}a. The plateau which is present for small
pseudorapidity
differences originates from the $\Delta R(\gamma,\ell)$ cut which
enforces a large azimuthal angle between the lepton and photon for
$|\Delta\eta(\gamma,\ell)|<0.7$, and hence reduces the cross section.
Since there is no radiation zero present in the dominating
$qg\rightarrow W\gamma q'$ and $\bar q'g\rightarrow W\gamma q$
processes, one does not expect a dip in the
1-jet $\Delta\eta(\gamma,\ell)$ distribution. Whereas inclusive NLO
QCD corrections almost completely fill in the dip at $\Delta\eta(\gamma,
\ell)=0$, a $\approx 20\%$ reduction of the differential cross section
remains in the NLO $W\gamma+0$~jet pseudorapidity difference
distribution (dotted line). For the
jet definition of Eq.~(\ref{EQ:LHCJET}), the Born and NLO
$W\gamma+0$~jet $\Delta\eta(\gamma,\ell)$ distributions almost coincide
in the region $|\Delta\eta(\gamma,\ell)|>1.5$ (see Fig.~\ref{FIG:ELEVEN}b).
Figure~\ref{FIG:ELEVEN} also demonstrates that the effect of higher order
QCD corrections on the pseudorapidity difference distribution can be
significantly reduced if a jet veto is imposed, {\it i.e.} if the
$W\gamma+0$~jet exclusive channel is considered. Nevertheless, NLO QCD
corrections to $W\gamma+0$~jet production significantly wash out the dip
signaling the radiation zero at the LHC. Non-standard $WW\gamma$
couplings do not affect the shape of the pseudorapidity difference
distribution significantly at the LHC when NLO QCD corrections are
included.

The result of the NLO $W\gamma+0$~jet
$\Delta\eta(\gamma,\ell)$ spectrum depends explicitly on the jet
definition used. Present studies~\cite{ATLAS,CMS} suggest that jets with
$p_T>50$~GeV can be identified at the LHC without problems, whereas it
will be very difficult to reconstruct jets with a transverse momentum
smaller than about 30~GeV. Even if the jet defining $p_T$ threshold is
lowered to 30~GeV, the dip in the NLO $W\gamma+0$~jet
$\Delta\eta(\gamma,\ell)$ distribution is not significantly
strengthened.

At energies higher than the LHC center of mass energy, the prospects
for observing the
SM radiation zero in the $\Delta\eta(\gamma,\ell)$ distribution rapidly
diminish. This is illustrated in Fig.~\ref{FIG:TWELVE}, where we
show the NLO pseudorapidity difference distribution for $pp\rightarrow
W^+\gamma+X$ at $\sqrt{s}=40$~TeV for the cuts summarized in
Eq.~(\ref{EQ:LHCCUTS}). Whereas the NLO $W\gamma+0$~jet
$\Delta\eta(\gamma,\ell)$ distribution at LHC energies still shows a
slight dip, there is essentially no trace left for $pp$ collisions at
$\sqrt{s}=40$~TeV (dotted line).

\section{SUMMARY AND CONCLUSIONS}

We have considered lepton photon pseudorapidity correlations as a tool to
study the radiation zero predicted by the SM for $W\gamma$ production
in hadronic collisions. In the SM, the dominant $W^\pm$ helicity in
$W\gamma$ production is $\lambda_W=\pm 1$. Combined with the $V-A$ coupling of
the charged lepton to the $W$, this implies that the lepton tends to be
emitted in the direction of the parent $W$, thus inheriting most of its
kinematic properties. As a result we found that, at the Tevatron,
the SM radiation zero
leads to a pronounced valley in the double differential distribution
$d^2\sigma/d\eta(\gamma)d\eta(\ell)$ for $W^\pm\gamma$ production and
rapidities fulfilling the relation $\Delta\eta(\gamma,\ell)=
\eta(\gamma)-\eta(\ell)\approx \mp 0.3$.

The pseudorapidity
difference distribution, $d\sigma/d\Delta\eta(\gamma,\ell)$, can be
studied as an alternative to the double differential distribution. Here
the radiation zero is signaled by a dip located at
$\Delta\eta(\gamma,\ell)\approx\mp 0.3$. The details of the rapidity
difference distribution are sensitive to the cuts imposed;
increasing the photon $p_T$ threshold and reducing the photon lepton
isolation cut tends to fill in the dip. A similar trend is observed for
non-standard $WW\gamma$ couplings, and if
NLO QCD corrections are taken into account. However, if the a jet veto is
imposed, {\it i.e.} the exclusive $W\gamma+0$~jet channel is considered,
the NLO rapidity difference distribution at the Tevatron
is very similar to that obtained in the Born approximation. The
influence of the $W+$~jet background, where the jet fakes a photon,
on the dip signaling the SM radiation zero strongly depends on the photon jet
misidentification probability, $P_{\gamma /j}$, which is detector
specific. If $P_{\gamma /j}$ drops sharply with the jet transverse energy, as
observed by UA2~\cite{UA2} and CDF, the $W+$~jet background was found not to
affect the significance of the dip in a noticeable way. For a (almost)
constant misidentification probability, which is consistent with the
result obtained by D\O~\cite{TONY}, the dip is partially filled in if
the background is not subtracted.
Therefore, for a clean observation of the SM radiation zero
\begin{itemize}
\item a cluster transverse mass cut of $M_T(\ell\gamma;p\llap/_T)
>90$~GeV,
\item a large photon lepton separation cut of $\Delta R(\gamma,\ell)>0.7$,
\item a good photon and lepton pseudorapidity coverage,
\item a low photon transverse momentum cut
\item and a careful subtraction of the $W+$~jet background
\end{itemize}
are essential. Many
experimental and theoretical uncertainties can be reduced substantially
by normalizing the pseudorapidity
difference distribution to the inclusive $W$ cross section.

In $pp$ collisions, the location of the valley (dip) in
$d^2\sigma/d\eta(\gamma)d\eta(\ell)$ ($d\sigma/d\Delta\eta(\gamma,\ell)
$) is shifted to $\Delta\eta(\gamma,\ell)=0$.

Compared to the photon rapidity distribution in the parton center of
mass frame,
$d\sigma/dy^*(\gamma)$, the pseudorapidity difference distribution
has the advantage of being independent of the twofold ambiguity in the
reconstruction of the parton center of mass frame, which considerably
dilutes the dip signaling the radiation zero in the $y^*(\gamma)$
distribution. In contrast to the $W^\pm\gamma/Z\gamma$ cross
section ratio, which also reflects the radiation zero~\cite{BEO},
the $\Delta\eta(\gamma,\ell)$ distribution
does not depend on the $Z\gamma$ cross section, and the validity of the
SM for $p\bar p\rightarrow Z\gamma$.

At the Tevatron, NLO QCD corrections to $W\gamma$ production are modest
and do not significantly affect the dip in the $\Delta\eta(\gamma,\ell)$
distribution. At supercollider
energies, on the other hand, higher order QCD corrections are very
pronounced due to the very large $qg$ luminosity, and considerably
obscure the radiation zero, even if a jet veto is imposed. Given a
sufficiently large
integrated luminosity, experiments at the Tevatron studying lepton photon
pseudorapidity correlations thus offer a {\it unique} opportunity to
search for the SM radiation zero in hadronic $W\gamma$ production.

\acknowledgements

We would like to thank H.~Aihara, P.~Grannis, S.~Keller, A.~Spadafora and
D.~Zeppenfeld for useful and stimulating
discussions. This research was supported in part by the
U.~S.~Department of Energy under Grant No.~DE-FG02-91ER40677 and
Contract No.~DE-FG05-87ER40319, and by the National Science Foundation
under Award No.~PHY9120537.

%
%
\newpage
%
%

%
\newpage
%
%
\figure{The double differential distribution $d^2\sigma/dy(\gamma)
dy(W^+)$ for $p\bar p\rightarrow W^+\gamma\rightarrow
\ell^+p\llap/_T \gamma$, $\ell=e,\,\mu$, in the Born approximation at
the Tevatron ($\sqrt{s}=1.8$~TeV). The cuts imposed are described in the text.
\label{FIG:ONE} }
%
\figure{The $W$ lepton rapidity difference distribution,
$d\sigma/d\Delta y(W,\ell)$ for $p\bar p\rightarrow W^+\gamma\rightarrow
\ell^+p\llap/_T \gamma$, $\ell=e,\,\mu$, in the Born approximation at
the Tevatron. The solid line shows the result obtained for the cuts
summarized in Eqs.~(\ref{EQ:TEVCUTS}) and~(\ref{EQ:MTCUT}). The dashed
curve displays the rapidity difference
distribution if the $p_T(\ell)$ and $p\llap/_T$ cuts are removed.
\label{FIG:TWO} }
%
\figure{The double differential distribution $d^2\sigma/d\eta(\gamma)
d\eta(\ell^+)$, $\ell=e,\,\mu$, for a) $p\bar p\rightarrow
\ell^+p\llap/_T\gamma$ and b) $p\bar p\rightarrow
\ell^+\ell^-\gamma$ in the Born approximation at the Tevatron
($\sqrt{s}=1.8$~TeV). The cuts imposed are described in the text.
\label{FIG:THREE} }
%
\figure{The photon lepton pseudorapidity difference distribution,
$d\sigma/d\Delta\eta(\gamma,\ell^+)$ at the Tevatron for $p\bar
p\rightarrow\ell^+p\llap/_T\gamma$ (solid line) and $p\bar
p\rightarrow\ell^+\ell^-\gamma$ (dashed line) in the Born
approximation. The cuts imposed are summarized in Eqs.~(\ref{EQ:TEVCUTS}),
(\ref{EQ:MTCUT}) and~(\ref{EQ:ZCUTS}).
\label{FIG:FOUR} }
%
\figure{The photon lepton pseudorapidity difference distribution,
$d\sigma/d\Delta\eta(\gamma,\ell)$ at the Tevatron for $p\bar
p\rightarrow\ell^+p\llap/_T\gamma$ in the Born
approximation. The solid line shows the full distribution. The dotted
curve displays the contribution of the sea quarks in both the proton and
the antiproton. The dashed line gives the $\Delta\eta(\gamma,\ell)$
distribution assuming that valence quarks in the proton and antiproton
only contribute. The dash-dotted line finally represents the portion
originating from valence quarks colliding with sea quarks. The cuts
imposed are summarized in Eqs.~(\ref{EQ:TEVCUTS}) and~(\ref{EQ:MTCUT}).
\label{FIG:FIVE} }
%
\figure{The pseudorapidity difference distribution,
$d\sigma/d\Delta\eta(\gamma,\ell)$, for $p\bar p\rightarrow
W^+\gamma\rightarrow\ell^+p\llap/_T\gamma$,  $\ell=e,\,\mu$, in the
Born approximation.
a) The solid line shows the result obtained for the cuts described in the
text. The dashed curve displays the pseudorapidity difference distribution
if the $p_T(\gamma)>5$~GeV cut is replaced by $p_T(\gamma)>10$~GeV,
with all other cuts unchanged. The dash-dotted line shows the result if
the $p_T(\ell)$, $p\llap/_T>20$~GeV cut is removed.
b) The pseudorapidity difference distribution for the set of cuts listed in
Eqs.~(\ref{EQ:TEVCUTS}) and~(\ref{EQ:MTCUT}) (solid line) is compared with
the result obtained if the
$\Delta R(\gamma,\ell)>0.7$ cut is replaced by $\Delta R(\gamma,\ell)>
0.3$, with all other cuts unchanged (dotted curve). The dashed double
dotted curve shows the result if the photon and lepton pseudorapidity
cuts of Eq.~(\ref{EQ:TEVCUTS}) are replaced by $|\eta(\gamma)|$,
$|\eta(\ell)|<1$.
\label{FIG:SIX} }
%
\figure{A closeup of the region around the minimum in
$d\sigma/d\Delta\eta(\gamma,\ell)$ for $p\bar p\rightarrow
W^+\gamma\rightarrow\ell^+p\llap/_T\gamma$,  $\ell=e,\,\mu$.
The solid line shows the result obtained for the cuts described in the
text. The dashed (dotted) curve displays the pseudorapidity difference
distribution if the $p_T(\gamma)>5$~GeV ($\Delta R(\gamma,\ell)>0.7$) cut is
replaced by $p_T(\gamma)>10$~GeV ($\Delta R(\gamma,\ell)>0.3$),
with all other cuts unchanged. The dash-dotted line shows the result if
the $p_T(\ell)$, $p\llap/_T>20$~GeV cut is removed. The long-dashed
line, finally, shows the pseudorapidity difference distribution if the
$\Delta R(\gamma,\ell)>0.7$ requirement is relaxed to $\Delta R(\gamma,
\ell)>0.1$, with all other cuts remaining unchanged.
\label{FIG:SEVEN} }
%
\figure{The differential cross section for the photon lepton
pseudorapidity difference for $p \bar p \to W^+ \gamma \to \ell^+\nu
\gamma$ at $\sqrt{s} = 1.8$~TeV in the SM.
a) The inclusive NLO $W\gamma+X$ differential cross section (solid line) is
shown, together with the the (LO) $W\gamma+1$~jet (dashed line) exclusive
differential cross section, and the $\Delta\eta(\gamma,\ell)$
distribution in the Born approximation (dash-dotted line),
using the jet definition in Eq.~(\ref{EQ:TEVJET}).
b) The NLO $W\gamma +0$~jet exclusive differential cross section
(dotted line) is compared with the Born differential cross section
(dash-dotted line). The cuts imposed are described in the text.
\label{FIG:EIGHT} }
%
\figure{The pseudorapidity difference distribution,
$d\sigma/d\Delta\eta(\gamma,\ell)$, for $p\bar p\rightarrow
W^+\gamma\rightarrow\ell^+p\llap/_T\gamma$,  $\ell=e,\,\mu$, at
the Tevatron in the Born
approximation for anomalous $WW\gamma$ couplings. The curves are for
the SM (solid), $\Delta\kappa_0=2.6$
(dashed), and $\lambda_0=1.7$ (dotted). Only one coupling is varied at a
time. A dipole form factor ($n=2$) with scale
$\Lambda=1$~TeV (see Eq.~(\ref{EQ:KAPPAFORM})) is assumed for
non-standard $WW\gamma$ couplings. The
cuts imposed are described in the text. The error bars indicate the
expected statistical uncertainties for an integrated luminosity of
22~pb$^{-1}$.
\label{FIG:NINE} }
%
\figure{a) The pseudorapidity difference distribution,
$d\sigma/d\Delta\eta(\gamma,\ell)$, for the $W+$~jet fake background,
$p\bar p\rightarrow W^+j\rightarrow\ell^+p\llap/_T``\gamma$'',
$\ell=e,\,\mu$, at the Tevatron. The solid line shows the prediction
obtained with the transverse momentum dependent jet photon
misidentification probability, $P_{\gamma /j}$, obtained by CDF. The
dashed line gives the result for a constant misidentification
probability of $P_{\gamma /j}=1\cdot 10^{-3}$, which is consistent with
the value measured by D\O.
b) $d\sigma/d\Delta\eta(\gamma,\ell)$ for $W^+\gamma$ production at
the Tevatron. The dotted and
dashed lines show the expected distribution for signal plus ($W+$~jet)
background, using the misidentification probability obtained by CDF and
$P_{\gamma /j}=1\cdot 10^{-3}$, respectively. For comparison, the
prediction of the signal alone (solid line) is also given. The cuts
imposed are summarized in Eqs.~(\ref{EQ:TEVCUTS}) and~(\ref{EQ:MTCUT}).
\label{FIG:TEN} }
%
\figure{The differential cross section for the photon lepton
pseudorapidity difference for $pp \to W^+ \gamma \to \ell^+\nu \gamma$
at $\sqrt{s} = 14$~TeV in the SM.
a) The inclusive NLO differential cross section (solid line) is
shown, together with the ${\cal O}(\alpha_s)$ 0-jet (dotted line),
and the (LO) 1-jet (dashed line) exclusive differential cross sections,
using the jet definition of Eq.~(\ref{EQ:LHCJET}).
b) The NLO $W\gamma +0$~jet exclusive differential cross section
(dotted line) is compared with the Born differential cross section
(dash-dotted line). The cuts imposed are summarized in Eq.~(\ref{EQ:LHCCUTS}).
\label{FIG:ELEVEN} }
%
\figure{The differential cross section for the photon lepton
pseudorapidity difference for $pp \to W^+ \gamma \to \ell^+\nu \gamma$
at $\sqrt{s} = 40$~TeV in the SM.
a) The inclusive NLO differential cross section (solid line) is
shown, together with the ${\cal O}(\alpha_s)$ 0-jet (dotted line),
and the (LO) 1-jet (dashed line) exclusive differential cross sections,
using the jet definition of Eq.~(\ref{EQ:LHCJET}).
b) The NLO $W\gamma +0$~jet exclusive differential cross section
(dotted line) is compared with the Born differential cross section
(dash-dotted line). The cuts imposed are summarized in Eq.~(\ref{EQ:LHCCUTS}).
\label{FIG:TWELVE} }
%
%
\end{narrowtext}
\end{document}